\documentclass[prl,twocolumn,superscriptaddress,a4paper]{revtex4-1}
\usepackage{graphicx}
\usepackage{enumitem}
\usepackage{amsmath}

\usepackage{amsfonts}

\usepackage[colorlinks,breaklinks,citecolor=blue,linkcolor=black,urlcolor=black]{hyperref}

\begin{document}

\title{Quantum liquids and droplets with low-energy interactions in one dimension}

\author{Ivan Morera}
\affiliation{Departament de F{\'i}sica Qu{\`a}ntica i Astrof{\'i}sica, Facultat de F{\'i}sica, Universitat de Barcelona, E-08028 Barcelona, Spain}
\affiliation{Institut de Ci{\`e}ncies del Cosmos, Universitat de Barcelona, ICCUB, Mart{\'i} i Franqu{\`e}s 1, E-08028 Barcelona, Spain}
\author{Bruno Juli{\'a}-D{\'i}az}
\affiliation{Departament de F{\'i}sica Qu{\`a}ntica i Astrof{\'i}sica, Facultat de F{\'i}sica, Universitat de Barcelona, E-08028 Barcelona, Spain}
\affiliation{Institut de Ci{\`e}ncies del Cosmos, Universitat de Barcelona, ICCUB, Mart{\'i} i Franqu{\`e}s 1, E-08028 Barcelona, Spain}
\author{Manuel Valiente}
\email{mvaliente@tsinghua.edu.cn}
\affiliation{Institute for Advanced Study, Tsinghua University, Beijing 100084, China}

\begin{abstract}
We consider interacting one-dimensional bosons in the universal low-energy regime. The interactions consist of a combination of attractive and repulsive parts that can stabilize quantum gases, droplets and liquids. In particular, we study the role of effective three-body repulsion, in systems with weak attractive pairwise interactions. Its low-energy description is often argued to be equivalent to a model including only two-body interactions with non-zero range. Here, we show that, at zero temperature, the equations of state in both theories agree quantitatively at low densities for overall repulsion, in the gas phase, as can be inferred from the $S$-matrix formulation of statistical mechanics. However, this agreement is absent in the attractive regime, where universality only occurs in the long-distance properties of quantum droplets. We develop analytical tools to investigate the properties of the theory, and obtain astounding agreement with exact numerical calculations using the density-matrix renormalization group. 
\end{abstract}

\maketitle
{\bf Introduction.}
Recent advances in the preparation, manipulation and observation of ultracold atomic droplets and liquids \cite{Cabrera2018,Schmitt2016,Ferrier2016,Ferioli2019,Bottcher2019,Semeghini2018} have sparked renewed interest in the physics of these systems \cite{Cui2018,Hu2020a,Hu2020b,Ferioli2020,Astrakharchik2018a,Parisi2020,Wang2020,Bottcher2021,Morera2020,Morera2021,Cikojevic2018,Cikojevic2020a,Cikojevic2020b}. This is more so due to their low densities and temperatures, allowing universal low-energy descriptions that are independent of the short-distance details of the relevant interactions \cite{Bloch2008}. The possibility to effectively confine these systems to one spatial dimension \cite{Kinoshita2004,Paredes2004}, where interaction effects are enhanced \cite{Giamarchi}, makes ultracold atoms a promising platform to realize highly controllable strongly interacting droplets and liquids. With traditional quantum liquids such as $^4\mathrm{He}$, interatomic potentials that reproduce essentially all of their experimentally measurable properties are known accurately \cite{Kunitski2020}. In deep contrast, the underlying interactions in ultracold atomic systems are highly dependent on the particular atomic species, and applied external fields, such as those involved in magnetic Feshbach resonances \cite{Chin2010} and transversal confinement \cite{Olshanii1998}. Hence, it is impractical, if not impossible, to attempt as accurate a description as in $^4\mathrm{He}$ for each realization of an ultracold atomic liquid. Therefore, a universal low-energy description of these systems, within the effective field theory (EFT) paradigm \cite{Bedaque2002,Hammer2013}, is highly desirable. 

In the two-body sector, the simplest EFT including scattering length and both physical and effective ranges has been recently used to describe a one-dimensional dimerized liquid in an optical lattice \cite{Morera2021}. This system is typically claimed to be equivalent, at low energies, with a theoretically simpler EFT that includes the two-body scattering length and a three-body contact interaction \cite{Hammer2013,Bulgac2002}, the latter being an emergent property due to three-body processes with two-body interactions that occur off the energy shell \cite{Valiente2019a,Pricoupenko2020}. In this Letter, we investigate one-dimensional quantum liquids and droplets at zero temperature that are described by such low-energy theories. We find that for overall repulsion the equations of state for the two-body theory and the EFT including three-body interaction agree well with each other at low densities, in accordance with the $S$-matrix formulation of statistical mechanics \cite{Dashen1969}. However, for overall attraction their droplets and liquid phases are not equivalent. This fact is proven by studying universal long-distance asymptotics in both models. We also develop highly non-perturbative, analytical approximations whose predictions are in excellent agreement with exact calculations using the density-matrix renormalization group (DMRG).

{\bf Low-energy theory.}
Our first aim is to exactly simulate a one-dimensional many-body system 
of bosons with attractive two-body and repulsive 
three-body interactions in free space at zero temperature. Ideally, one could apply ground-state methods, 
such as ground state quantum Monte Carlo, to the low-energy Hamiltonian, 
given by~\cite{Pastukhov2019,Valiente2019a,Valiente2019b,Valiente2020a,Valiente2020b,Sekino2018,Hou2020,Czejdo2020,Hou2019,Drut2018,Pricoupenko2019,Pricoupenko2020,Guijarro2018,Nishida2018,LPricoupenko2018,LPricoupenko2018,LPricoupenko2019}
\begin{align}
  H_{c}&=\sum_{i=1}^{N}\frac{p_i^2}{2m}+g_0\sum_{i<j=1}^N\delta(x_i-x_j)\nonumber\\
  &+g_3\sum_{i<j<k=1}^N\delta(x_i-x_j)\delta(x_j-x_k),\label{Hamiltonian}
\end{align}
where $m$ is the mass of the particles, $g_0=-2\hbar^2/ma$ is the Lieb-Liniger coupling constant, with $a$ ($>0$) the scattering length, and $g_3$ is the 
bare three-body coupling constant~\cite{Pastukhov2019,Valiente2019a,Sekino2018}. 
With square cutoff ($\Lambda$) regularization, it is given 
by $\hbar^2/g_3(\Lambda)m=\ln|Q_*/\sqrt{3}\Lambda|/\sqrt{3}\pi$, with $Q_*$ 
the three-body momentum scale~\cite{Valiente2019a} beyond which the low-energy 
theory~(\ref{Hamiltonian}) breaks down. If the three-body interaction is repulsive and 
$g_0=0$, the $T-$matrix exhibits a Landau pole at 
energy $E=-\hbar^2Q_*^2/2m$, which prevents the physical $N$-body ($N>2$) ground 
state from being explored using ground state 
methods~\cite{Sekino2018,Nishida2018}. Below, we solve this issue
by discretizing the problem on a lattice near, but not in the continuum 
limit, with a genuinely repulsive three-body force. 


The simplest direct discretization of Hamiltonian (\ref{Hamiltonian}) is given 
by a generalized Bose-Hubbard Hamiltonian, 
\begin{align}
  H&=-J\sum_j(b_{j+1}^{\dagger}b_j+\mathrm{H.c.})+\frac{U_2}{2}\sum_j n_j(n_j-1)\nonumber\\
  &+\frac{W}{6}\sum_jn_j(n_j-1)(n_j-2)+2J\sum_jn_j.\label{HamiltonianLattice}
\end{align}
Above, $J=\hbar^2/(2md^2)$ is the hopping strength, with $d$ the lattice spacing, $U_2=g_0/d$ is 
the on-site two-body interaction strength, $W$ ($>0$) is a three-body 
coupling constant to be determined, $b_j$ ($b_j^{\dagger}$) 
annihilates (creates) a boson at site $j$ ($\in \mathbb{Z}$), 
and $n_j=b_j^{\dagger}b_j$ is the local number operator. To relate 
the three-body lattice coupling constant $W$ to the continuum 
momentum scale $Q_*$, we solve the three-body problem with $U_2=0$ 
at vanishing total quasimomentum, and match the continuum and 
lattice amplitudes at low energies \cite{Valiente2019a}, obtaining $W/J=(\beta+\ln|Q_*d|/2\sqrt{3})^{-1}$,
%
%
where $\beta=-0.1956\ldots$ is a numerical constant of no physical 
relevance, i.e., it is regularization-dependent. 
Hamiltonian (\ref{HamiltonianLattice}) can then 
be used to simulate a continuum repulsive three-body force at low 
densities with a finite lattice spacing, provided that $W/J>0$, i.e., 
for $|Q_*d|>\exp(2\sqrt{3}|\beta|)\approx 8.4$, corresponding to the 
weak to moderate coupling regime for the three-body 
interaction~\cite{Pastukhov2019,Valiente2019a,Valiente2019b,Valiente2020a,Valiente2020b,Sekino2018,Hou2020,Czejdo2020,Hou2019,Drut2018,Pricoupenko2019,Pricoupenko2020,Guijarro2018,Nishida2018,LPricoupenko2018,LPricoupenko2018,LPricoupenko2019}.

\begin{figure}[t]
\begin{center}
\includegraphics[width=0.49\textwidth]{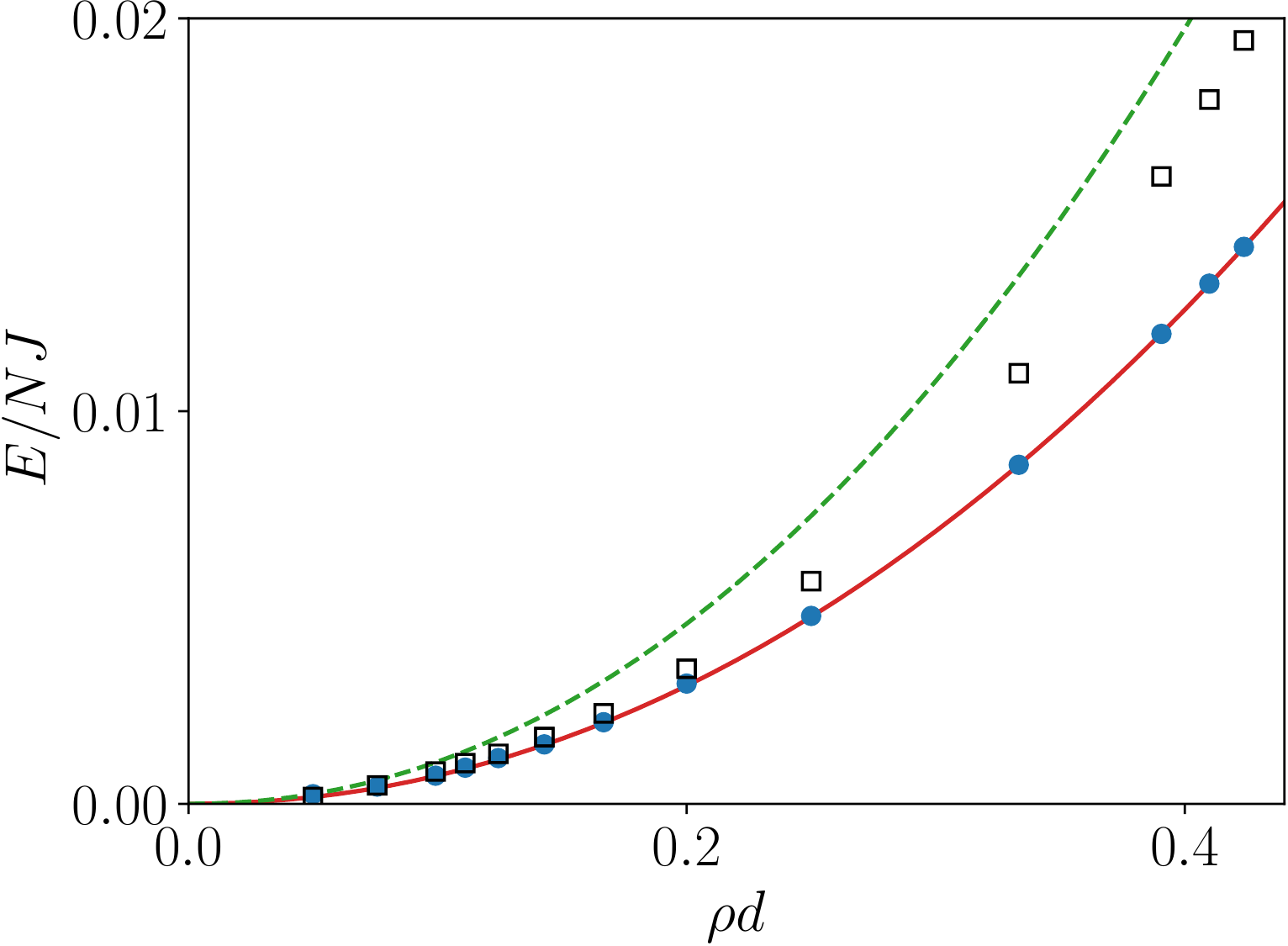}
\end{center}
\caption{Zero-temperature equation of state for 
Hamiltonian (\ref{HamiltonianLattice}) with $U_2=0$ and $W/J=1.1$ 
(filled blue dots), weak-coupling EoS, Eq.~(\ref{EoSWeak}), 
at scale $\mu=\exp(\gamma)Q_*^2/\sqrt{8}\rho$, corresponding to Pastukhov's scale \cite{Pastukhov2019} (green dashed line), 
and at the renormalization scale $\mu=188.464/\rho d^2$ (red solid line). Open black squares correspond 
to the EoS obtained from the extended Hubbard model at its two-body 
resonance (see text).} 
\label{fig:gas}
\end{figure}

To test the ability of the lattice Hamiltonian, Eq.~(\ref{HamiltonianLattice}), 
to describe the three-body repulsive side of its continuum counterpart, 
Eq.~(\ref{Hamiltonian}), we obtain the zero-temperature equation 
of state (EoS) of Hamiltonian (\ref{HamiltonianLattice}) with $U_2=0$ 
and $W/J=1.1$ using DMRG, and compare it with the  
weak-coupling expansion due to Pastukhov \cite{Pastukhov2019}
\begin{align}
  \frac{E}{N}&=\frac{\hbar^2\rho^2}{3!m}g(\mu)\Bigg[ 1-\frac{4}{\pi}[g(\mu)]^{1/2} \nonumber\\ &+\frac{g(\mu)}{\sqrt{3}\pi}\left\{\frac{1}{2}\ln\left|\frac{\sqrt{3}\pi}{g(\mu)}\right|-C\right\}\Bigg],\label{EoSWeak}
\end{align}
where $\mu=\mu(\rho)=\xi/\rho$ is a renormalization scale, $g(\mu)=-\sqrt{3}\pi/\ln|Q_*/\mu|$ 
is the renormalized coupling constant~\cite{Pastukhov2019}, 
and $C=-9.863\ldots$. The ambiguity, in perturbation theory, to choose 
the scale $\mu$ is identical to that in the 2D Bose 
gas~\cite{Beane2010,Beane2018}. To obtain predictive power, we match 
the weak-coupling EoS (\ref{EoSWeak}) at only one value of the density 
$\rho$ with the EoS at that density obtained with DMRG. The results are 
shown in Fig.~\ref{fig:gas}, where astounding agreement is observed 
over all ranges of density.

{\bf Quantum liquid.}
We now turn our attention to the ground state liquid phase of 
Hamiltonian (\ref{Hamiltonian}), and study it both theoretically, 
with two different approximations, and numerically using its 
lattice discretization~(\ref{HamiltonianLattice}). We begin by introducing 
a non-perturbative decoupling approximation for the three-body 
interaction of Hamiltonian (\ref{Hamiltonian}), valid for large 
particle numbers ($N\to\infty$), while treating the two-body 
interaction exactly. In this approximation, the Hamiltonian $H_{D}$ 
takes the form~\cite{Supplemental}
\begin{equation}
  H_{D}=\sum_{i=1}^N\frac{p_i^2}{2m}+G_N(\Gamma)\sum_{i<j=1}^{N}\delta(x_i-x_j)+C_N(\Gamma),\label{HamiltonianD}
\end{equation}
which, up to a constant, has the form of the Lieb-Liniger model 
with coupling constant $G_N(\Gamma)$. In Eq.~(\ref{HamiltonianD}), 
$\Gamma=\langle \delta(x_1-x_2) \rangle$, $G_N(\Gamma)=g_0+(N-2)g_3\Gamma$, 
and $C_N=-\Gamma^2g_3N(N-1)(N-2)/3$. In this approximation, $g_3$ ceases to be a bare coupling 
constant (i.e. cutoff-dependent), and instead becomes a renormalized 
one (i.e. finite) to be fixed to reproduce certain physical properties, 
such as the equilibrium binding energy. We note that, since the original 
Hamiltonian~(\ref{Hamiltonian}) contains a logarithmic anomaly upon 
renormalization, the coupling constant $g_3$ in this approximation 
depends on the state. In particular, in the ground state, it depends 
on the bulk density of the droplet, or the particle number. For 
$N\to\infty$, and in free space ($L=\infty$), that is, at equilibrium, 
the ground state $\psi_N$ and its energy $E_0(N)$ can be obtained 
exactly using MacGuire's solution~\cite{MacGuire1966}. Solving the 
problem self-consistently~\cite{Supplemental}, we obtain
\begin{equation}
  E_0(N)\to -\frac{1}{3}\frac{g_0^2}{g_3^{\mathrm{eq}}}N,\hspace{0.1cm} N\to\infty \,,\label{E0bulk}
\end{equation}
%
\begin{figure}[t]
\begin{center}
\includegraphics[width=0.48\textwidth]{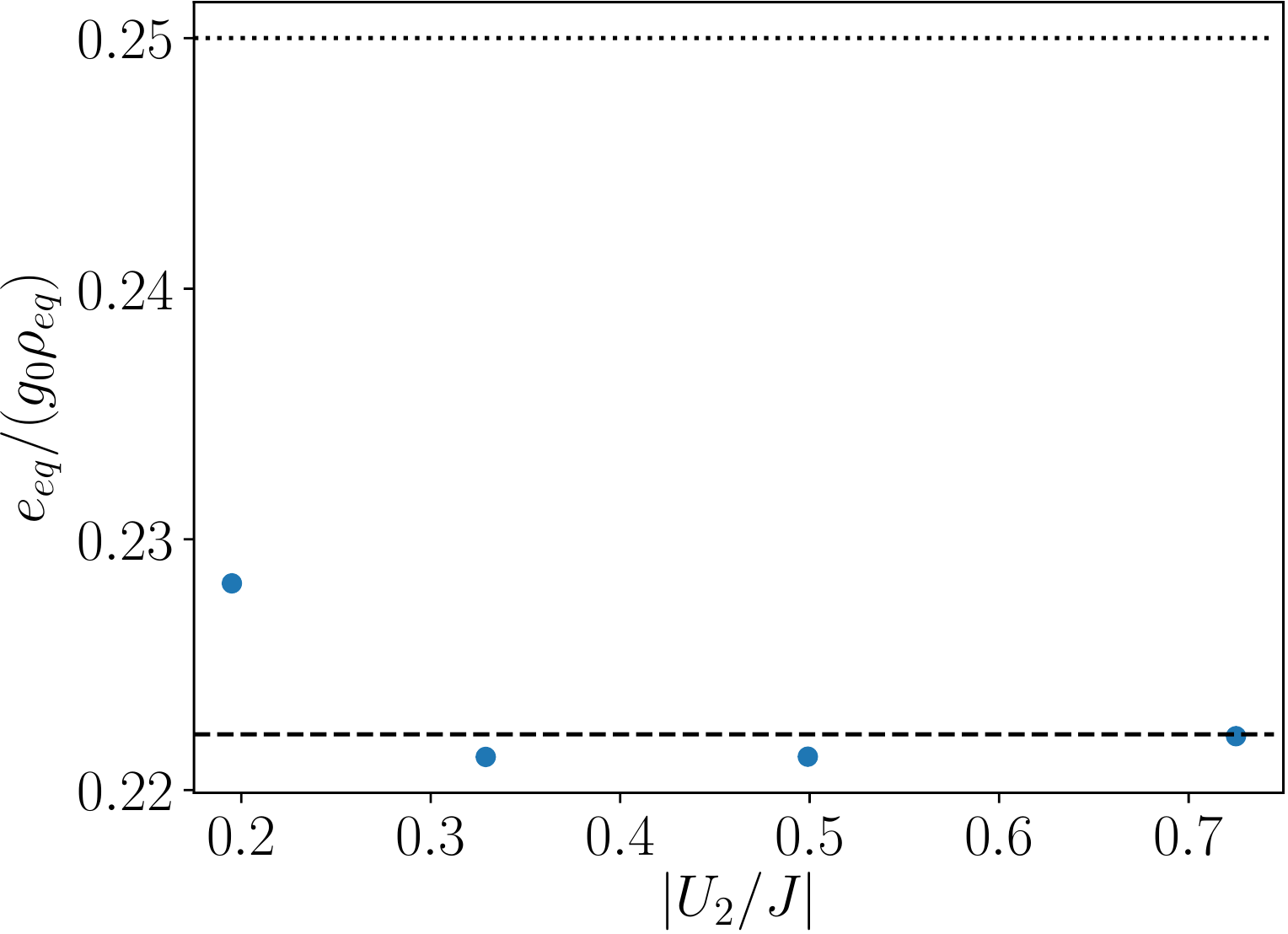}
\end{center}
\caption{Ratio $e_0/g_0\rho_{\mathrm{eq}}$ calculated numerically (blue dots) 
for Hamiltonian (\ref{HamiltonianLattice}) as function of two-body 
interaction strength. The three-body interaction strengths for points 
labelled with $U_2/J=-0.195$, $-0.329$, $-0.499$ and $-0.725$ are 
given by $W/J=1.28$, $1.4$, $1.48625$ and $1.508$, respectively. Dashed line ($2/9=0.222\ldots$) is the prediction from Eq.~(\ref{Erhoeqlinear}), and dotted line is the standard mean field prediction.} 
\label{fig:Constant}
\end{figure}
%
where $g_3^{\mathrm{eq}}$ is the value of $g_3$ at equilibrium. The 
density profile $\rho_N(x)$ can be obtained by computing the density 
with respect to a fixed center-of-mass coordinate $X$ as~\cite{Calogero1975}
\begin{equation}
  \rho_N(x) = \int \mathrm{d}x_2\ldots\mathrm{d}x_N\delta(X)|\psi_N(x,x_2,\ldots,x_N)|^2\,.\label{rhoN}
\end{equation}
Its bulk density $\rho_{\mathrm{eq}}$, given 
by $\rho_{\mathrm{eq}}=\lim_{N\to\infty}\rho_N(0)$, is known 
analytically for $N\to\infty$~\cite{Calogero1975,Gertjerenken2012} and, as opposed to the 
soliton of the usual Lieb-Liniger model with no three-body repulsion, 
it is finite because $G_N=O(N^{-2})$ for large $N$. We have
\begin{equation}
  \rho_N(0)=-\frac{1}{4}\frac{mG_N}{\hbar^2}N(N-1)\to \frac{3}{2}\frac{|g_0|}{g_3^{\mathrm{eq}}}=\rho_{\mathrm{eq}}\,.\label{rhobulk}
\end{equation}
Defining $\mathrm{e}_{\mathrm{eq}}=\lim_{N\to\infty}E_0/N$, Eqs.~(\ref{E0bulk}) 
and~(\ref{rhobulk}) can be combined to eliminate the (unknown) effective 
three-body coupling $g_3^{\mathrm{eq}}$ in favour of measurable physical 
quantities, obtaining
\begin{equation}
  e_{\mathrm{eq}}=\frac{2}{9}g_0\rho_{\mathrm{eq}}\,.\label{Erhoeqlinear}
\end{equation}
The above equation, which is approximate, yet highly non-perturbative, 
is one of the main results of this Letter. It predicts a strongly 
constrained, linear relation between the equilibrium energy per particle 
and density for fixed two-body interaction strength $g_0<0$. To 
test Eq.~(\ref{Erhoeqlinear}), in Fig.~\ref{fig:Constant}, we plot the 
relation $e_{\mathrm{eq}}/g_0\rho_{\mathrm{eq}}$ calculated for 
Hamiltonian (\ref{HamiltonianLattice}) using DMRG for a number of 
different values of the pair $(U_2,W)$. The results show that not 
only is the linear relation in Eq.~(\ref{Erhoeqlinear}) a very 
good approximation, but also the prediction of the proportionality constant $2/9=0.222\ldots$ is in excellent agreement with the exact results. Note 
that mean-field theory, with chemical potential $\mu=g_0\rho+g_3\rho^2/2$, 
also predicts a constant value for 
$e_{\mathrm{eq}}/|g_0|\rho_{\mathrm{eq}}=1/4$, which is $\sim 15-20\%$ 
off the numerical values and the prediction of Eq.~(\ref{Erhoeqlinear}).

\begin{figure}[t]
\begin{center}
\includegraphics[width=0.48\textwidth]{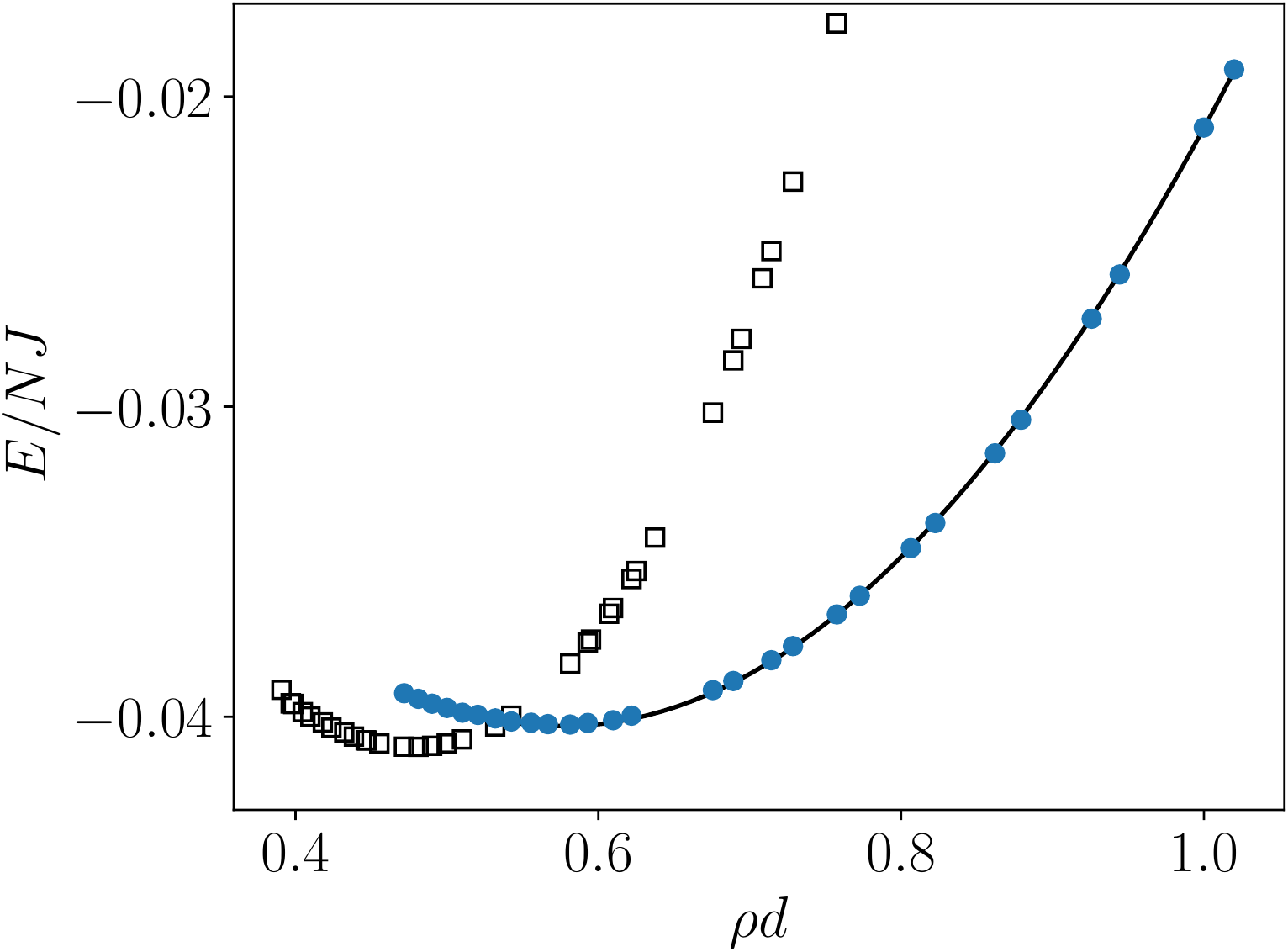}
\end{center}
\caption{Exact zero-temperature equation of state for 
Hamiltonian (\ref{HamiltonianLattice}) with $U_2/J=-0.33$ and 
$W/J=1.39$ (blue dots), and improved mean-field approximation, 
Eq.~(\ref{eIMF}), with 
$\rho_{\mathrm{eq}}d=0.576$, $mg_3^{\mathrm{eq}}/\hbar^2=0.9899$ and 
$\eta=0.2119$ (see text). Black open squares are the exact EoS of 
the extended Hubbard model with $U/J=10$ and $V/J=-8/5$ (see text).} 
\label{fig:iMF}
\end{figure}

The second approximation we introduce is an improved version of mean-field 
theory (iMF), by allowing for $g_3$ to depend logarithmically on the density as
$g_3(\rho)=g_3^{\mathrm{eq}}/[\lambda \ln|\rho/\rho_{\mathrm{eq}}|+1]$, 
so that it can account for the anomaly non-perturbatively.
The value $g_3^{\mathrm{eq}}$ is given by the coupling constant at 
equilibrium, and is fixed by $e_{\mathrm{eq}}$, while $\lambda$ is 
a dimensionless parameter that is fixed by the equilibrium 
density $\rho_{\mathrm{eq}}$, given $e_{\mathrm{eq}}$.
To lowest order, integrating the chemical potential gives for 
the energy per particle $e_{\mathrm{iMF}}$ in 
this approximation,
\begin{equation}
  e_{\mathrm{iMF}}(\rho)\approx\frac{1}{2}g_0\rho
  +\frac{1}{6}g_3(\rho)\rho^2\,.\label{eIMF}
\end{equation}
Given the freedom of scale in logarithmic running of the coupling 
constant, the above approximation is sufficient.
The relation between the equilibrium energy and density can also be 
obtained eliminating $g_3^{\mathrm{eq}}$ and is given by $e_{\mathrm{iMF}}(\rho_{\mathrm{eq}})=\eta g_0\rho_{\mathrm{eq}}$, with $\eta = (1/2)(1-\lambda)/(2-\lambda)$. The parameter $\lambda$ is universal in the decoupling approximation (see Eq.~(\ref{Erhoeqlinear})), and given by $\lambda=1/5$ ($\eta=2/9$), leaving effectively the iMF with only one free parameter, $g_3$. In Fig.~\ref{fig:iMF}, we plot the EoS 
at zero temperature for Hamiltonian (\ref{HamiltonianLattice}), together 
with the improved mean-field approximation, Eq.~(\ref{eIMF}), showing astounding agreement.

{\bf Universal and non-universal properties.}
Three-body interactions are always present in a many-atom system, and can 
be genuine or emergent from the off-shell structure of the two-body 
interactions~\cite{Valiente2019a,Pricoupenko2020,Hammer2013}. In general, 
these are a combination of genuine and emergent. For repulsive 
one-dimensional systems, two-body effective range (on-shell) effects 
are typically negligible for low densities~\cite{Astrakharchik2010}. 
However, the effects of a non-zero physical range (off-shell) are 
important for three-body processes, especially for large scattering 
lengths~\cite{Valiente2019a,Pricoupenko2020}, which can be rigorously 
modelled by simple three-body forces. This can also be explained 
qualitatively using field redefinitions~\cite{Hammer2013} that trade 
off-shell low-energy interactions in favour of simpler, on-shell 
three-particle forces. Since, in 
this case, the low-energy two- and three-body amplitudes are 
essentially identical, the $S$-matrix formulation of statistical 
mechanics~\cite{Dashen1969} implies thermodynamic equivalence in 
the gas phase which, at zero temperature, means this is the 
case for (overall) repulsive interactions. For attraction, where 
liquids may be formed, the results of Ref.~\cite{Dashen1969} do 
not apply and, as we shall see, there is no such equivalence.    

To investigate the claims discussed just above, we choose the simplest lattice Hamiltonian with large scattering length and non-zero effective range, given by the extended Hubbard model (EHM), corresponding to Hamiltonian (\ref{HamiltonianLattice}), with $U_2$ replaced by $U$, $W=0$, plus an interaction term of the form $\mathcal{V}=V\sum_jn_jn_{j+1}$. On the two-body resonance ($1/a=0$), matching the low-energy two-body amplitude for Hamiltonian (\ref{HamiltonianLattice}) requires $U_2=0$, while the EHM requires $U/J=-4(V/J)/(2+V/J)$ \cite{Valiente2009,Morera2021}. To fit the low-energy three-body amplitude for the EHM, it is simplest to consider the finite-size ground state energy of Hamiltonian (\ref{HamiltonianLattice}) with three particles, and match the corresponding energy for the EHM \cite{Luscher1986-1,Valiente2019a}. For the case of Fig.~\ref{fig:gas}, $W/J=1.1$, corresponding to $U/J=10$ and $V/J=-8/5$ in the EHM. In Fig.~\ref{fig:gas}, we observe very good agreement between the two EoS for relatively low densities, while for larger densities two-body on-shell effects appear to dominate in the EHM. 

In the attractive case, which features $N$-body bound states for all $N$, it is possible to fix the locations of the poles of the $S$-matrix (bound state energies) rather accurately in both models, but not the residues at the poles without further parametrization. These are related to the so-called asymptotic normalization coefficient (ANC) $\gamma_N$ for a bound state \cite{Taylor2000,Luscher1986-2,Koenig2012}. For two particles in one dimension, defining the normalized relative bound state $\psi_2(x)$, $\gamma_2$ is defined as $|\gamma_2|=\lim_{x_{12}\to\infty}\left[|\psi_2(x_{12})|\exp\left(\sqrt{mE_{2}^{(B)}/\hbar^2}|x_{12}|\right)\right]$, where $E_2^{(B)}$ ($>0$) is the binding energy. For large and positive (attractive) scattering length $a$ in comparison with the effective range $r$ ($a/|r|\gg 1$), the two-body binding energies obtained with and without including the effective range agree well with each other. The ANCs are also in rather good agreement. For example, for the two models considered here, we have $\left|\gamma_2^{\mathrm{E}}/\gamma_2\right|^2\approx (r_e/d) \left( 2-r_e/d\right)+\left(2-(r_e/d)^2 \right)/(a/d)$
where $\gamma_2^{\mathrm{E}}$ and $\gamma_2$ are, respectively, the two-body ANC of the EHM and Hamiltonian (\ref{HamiltonianLattice}) with identical scattering lengths. For the case studied in Fig.~\ref{fig:iMF}, we have $|\gamma_2^{\mathrm{E}}/\gamma_2|^2\approx 1.07$, which shows a small yet non-negligible deviation from unity.

For large particle numbers, we can show that the ratios $|\gamma_N^{\mathrm{E}}/\gamma_N|$ become exponential in $N$. Using the universal asymptotics in one dimension \cite{Koenig2012}, we obtain \cite{Supplemental} for $N$ bosons, as $|x|\to\infty$,
\begin{equation}
   \rho_N(x)\to \frac{\pi}{2}|\gamma_{N}|^2A_{N-1}\frac{N}{N-1}e^{-\frac{N}{N-1}2\kappa_{1,N}|x|}.\label{rhoasymptotic}
\end{equation}
Above, $\gamma_{N}$ is the ANC in the $N\leftrightarrow (N-1)+1$ breakup channel, $\hbar^2\kappa_{1,N}^2/2m = |E_N-E_{N-1}|(N-1)/N$ represents the binding energy with respect to the ground state with one fewer particle, and $A_N$ is a model independent normalization factor. Using Eq.~(\ref{rhoasymptotic}) for the two models here considered, we obtain, as $|x|\to\infty$, $\rho^{\mathrm{E}}_N(x)/\rho_N(x)\to\left|\gamma_{N}^{\mathrm{E}}/\gamma_{N}\right|^2$,
where the superscript E (no superscript) corresponds to the EHM (Hamiltonian (\ref{HamiltonianLattice})). This relation has important consequences. We assume that the asymptotic normalization coefficients for the two models are different for all $N$, since they already are for two particles. If an $N$-body bound state is a quantum droplet, then most of the particle content lies within the bulk of the droplet, with a density we may consider constant. The radius $R$ of the droplet is then $R\approx N/2\rho_N(0)$. One may approximate the exponential tails for large $N$ as $\rho_N(x)\sim \rho_N(0)\exp(-2\kappa_{1,N}|x-R|)$ as $|x|\to\infty$, obtaining
\begin{equation}
  \left|\frac{\gamma_{N}^{\mathrm{E}}}{\gamma_{N}}\right|^2\sim \frac{\rho_N^{\mathrm{E}}(0)}{\rho_N(0)}\exp\left[\kappa_{1,N}N\left(\frac{1}{\rho_N^{\mathrm{E}}(0)}-\frac{1}{\rho_N(0)}\right)   \right].\label{gammaexp}
\end{equation}
The above relation indicates that we require algebraically small ($O(N^{-1/2-|\epsilon|})$) differences between the ANCs in order to achieve algebraically ($O(N^{-1-|\delta|})$) small differences in the equilibrium densities of quantum droplets and liquids, and vice versa. A finite difference in the large-$N$ densities of the two models immediately implies a exponential disagreement between the ANCs, and therefore the residues at the poles of the $S$-matrix. Equation (\ref{gammaexp}) shows our claim that, in order for two different models of a quantum liquid to share thermodynamical properties, not only the bound state energies but also the residues at these energies must be matched, and that the matching must have algebraic precision. In Fig.~\ref{fig:iMF}, we plot the EoS for the EHM with $U/J=10$ and $V/J=-8/5$, for which we have verified that the binding energies for all $N$ match those of Hamiltonian (\ref{HamiltonianLattice}) with $U_2=-0.33$ and $W/J=1.39$ within $2\%$. It is observed that the equilibrium densities are in clear disagreement, and so are the EoS at all densities, as we expected. 

%
\begin{figure}[t]
\begin{center}
\includegraphics[width=0.48\textwidth]{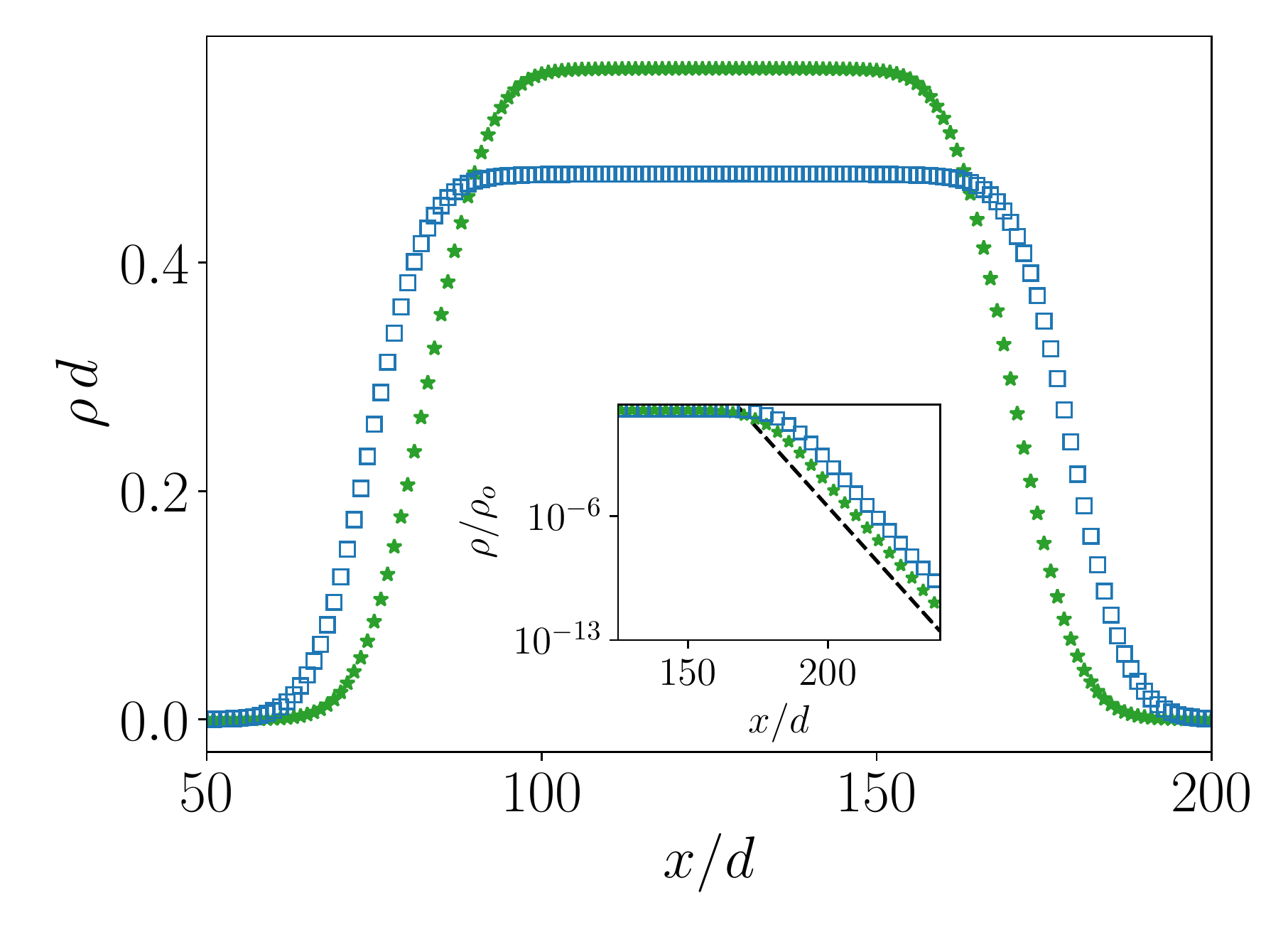}
\end{center}
\caption{Main panel: Density profiles with $N=50$ particles for Hamiltonian~\eqref{HamiltonianLattice} with $U_2/J=-0.33$ and $W/J=1.39$ (open squares) and the extended BH (see main text) with $U/J=10$ and $V/J=-1.6$ (stars). Inset panel: Density tails, rescaled by their respective saturation densities. Dashed line indicates the analytical result Eq.~\eqref{rhoasymptotic}. } 
\label{fig:droplet}
\end{figure}
The disagreement in the EoS between the two models has strong consequences when looking at the density profiles of the respective quantum droplets. Even though both models exhibit the same equilibrium energy, their equilibrium densities, and thus the droplet saturation densities, are different. This implies that droplets in the two models with equal particle number have different sizes, see Fig.~\ref{fig:droplet}, a direct consequence of the ANCs $\gamma_N$, Eq.~\eqref{gammaexp}. On the other hand, the decay of the density far away from the center of the droplet has to be identical in both models if the equilibrium energy is the same as predicted by Eq.~\eqref{rhoasymptotic}. In the inset panel of Fig.~\ref{fig:droplet} we show that both droplets exhibit the same exponential density decay far away from the respective centers. Moreover the decay is dictated by Eq.~\eqref{rhoasymptotic} which shows that it is directly related to the chemical potential at equilibrium for large number of particles.
Therefore, the tails of quantum droplets are universal for different models with identical binding energies (in free space), as opposed to the saturation density.

{\bf Conclusions.}
We have investigated one-dimensional quantum liquids at zero temperature and droplets using universal low-energy theories. We have shown that, while in the repulsive case different models with (almost) identical two- and three-body scattering amplitudes have identical low-density equations of state, small deviations in densities in the few-particle sector grow exponentially in the many-body limit. This implies the lack of equivalence of the zero-temperature equations of state. We have developed theoretical techniques that yield quantitatively accurate predictions in all density regimes, when compared to exact results obtained with DMRG. 

{\bf Acknowledgements}
This work has been partially supported by MINECO
(Spain) Grant No.FIS2017-87534-P. We acknowledge financial support from Secretaria d’Universitats i Recerca del Departament d’Empresa i Coneixement de la Generalitat de Catalunya, co-funded by the European Union
Regional Development Fund within the ERDF Operational Program of Catalunya (project QuantumCat, ref. 001-P-001644).

\pagebreak
\onecolumngrid
\vspace{\columnsep}
\newpage
\begin{center}
\textbf{\large Supplemental Material: Quantum liquids and droplets with low-energy interactions in one dimension}
\end{center}
\vspace{2cm}
\twocolumngrid

\setcounter{equation}{0}
\setcounter{figure}{0}
\setcounter{page}{1}
\makeatletter
\renewcommand{\theequation}{S\arabic{equation}}
\renewcommand{\bibnumfmt}[1]{[S#1]}
\renewcommand{\citenumfont}[1]{S#1}
\addtolength{\textfloatsep}{5mm}

\section{Decoupling approximation}
Here, we derive the approximate Hamiltonian (\ref{HamiltonianD}) by using a decoupling fluctuation expansion. This consists of approximating the product $\delta(x_i-x_j)\delta(x_j-x_k)$ as
\begin{align}
  \delta(x_i-x_j)\delta(x_j-x_k)&\approx \langle \delta(x_i-x_j) \rangle \delta(x_j-x_k) \nonumber \\
  &+ \langle \delta(x_i-x_k)\rangle \delta(x_i-x_j) \nonumber\\
  &+ \langle \delta(x_j-x_k)\rangle \delta(x_i-x_k)\nonumber\\
  &-2\langle \delta(x_i-x_j)\rangle \langle \delta(x_j-x_k)\rangle.
\end{align}
For identical bosons, we have $\Gamma\equiv \langle \delta(x_i-x_j)\rangle=\langle\delta(x_{i'}-x_{j'})\rangle$ for all $i\ne j$ and $i'\ne j'$. Hamiltonian (\ref{Hamiltonian}) reduces immediately to Hamiltonian (\ref{HamiltonianD}).

The ground state, for attractive interactions ($G_N(\Gamma)<0$), gives for the energy $E_0(N)$ \cite{MacGuire1966S}
\begin{equation}
  E_0(N) = -\frac{m\left[G_N(\Gamma)\right]^2}{24\hbar^2}N(N+1)(N-1)+C_N(\Gamma).\label{E0N-S}
\end{equation}
Using the Hellmann-Feynman theorem, and eliminating $\Gamma$, we obtain
\begin{align}
  \Gamma&=-\frac{mG_N}{6\hbar^2}(N+1),\label{Gamma-S}\\
  G_N&=-|g_0|\left[1-\frac{1}{1+\frac{6\hbar^2}{mg_3}[(N+1)(N-2)]^{-1}}\right].\label{GN-S}
\end{align}
Relation (\ref{Erhoeqlinear}) in the limit $N\to\infty$ is obtained by inserting Eqs.~(\ref{Gamma-S}) and (\ref{GN-S}) into Eq.~(\ref{E0N-S}) and taking the limit $N\to\infty$.

\section{Universal asymptotics}
Here, we derive Eq.~(\ref{rhoasymptotic}) for the asymptotic behaviour of the density profile of quantum droplets at large distances from the center. We begin by using the results of Ref.~\cite{Koenig2012S}, particularized to one spatial dimension. When one of the particles (e.g. particle 1) is asymptotically far from the rest of the system, the ground state wave function $\psi_N$ factorizes as
\begin{align}
  \psi_N&\to \gamma_{N}\sqrt{\kappa_{1,N}r_{1,N-1}}K_{-1/2}(\kappa_{1,N}r_{1,N-1})\nonumber\\
  &\times \psi_{N-1}(x_2,\ldots,x_N).\label{psiasymptotic-S}
\end{align}
Above, $\gamma_{N}$ is the asymptotic normalization coefficient in the $N\leftrightarrow (N-1)+1$ breakup channel, $\hbar^2\kappa_{1,N}^2/2m = |E_N-E_{N-1}|(N-1)/N$ represents the binding energy with respect to the ground state with one fewer particle, $K_{\nu}$ is the modified Bessel function of the second kind of order $\nu$, $r_{1,N-1}=|x_1-X_{N-1}|$, with $X_{N-1}=\sum_{i=2}^Nx_i/(N-1)$ the center of mass coordinate of the $(N-1)$-body subsystem, and $\psi_{N-1}$ is its ground state wave function. The density profile $\rho_N(x)$ of the bound state is given by Eq.~(\ref{rhoN}). At long distance away from the bulk of the droplet, we can introduce Eq.~(\ref{psiasymptotic-S}) into Eq.~(\ref{rhoN}), obtaining, as $|x|\to\infty$,
\begin{equation}
  \rho_N(x)\to \frac{\pi}{2}|\gamma_{1,N}|^2A_{N-1}\frac{N}{N-1}e^{-\frac{N}{N-1}2\kappa_{1,N}|x|},\label{rhoasymptotic-S}
\end{equation}
where $A_{N-1}=\int \mathrm{d}^{N-2}z|\psi_{N-1}|^2$ is a model-independent normalization factor. Collectively, $\mathbf{z}$ is an $(N-2)$-dimensional vector containing all degrees of freedom of the $(N-1)$-particle system except for the center-of-mass coordinate.

 \bibliographystyle{unsrt}

\end{document}